\documentstyle[preprint,tighten,aps]{revtex}

\def\lapprox{\mathrel{\mathop
  {\hbox{\lower0.5ex\hbox{$\sim$}\kern-0.8em\lower-0.7ex\hbox{$<$}}}}}
\def\gapprox{\mathrel{\mathop
  {\hbox{\lower0.5ex\hbox{$\sim$}\kern-0.8em\lower-0.7ex\hbox{$>$}}}}}

\def\N_F{$n_F$}

\begin{document}

\preprint{\vbox{\noindent
          \null\hfill  INFNFE-15-98}}
\title{Bounds on  $hep$ neutrinos}

\author{ G.~Fiorentini$^{1,2,3}$,
         V. Berezinsky$^{4}$,
         S.~Degl'Innocenti$^{5,2}$
         and B.~Ricci$^{1,2}$
       }
\address{
$^{1}$Dipartimento di Fisica dell'Universit\`a di Ferrara, 
       via Paradiso 12, I-44100 Ferrara, Italy
$^{2}$Istituto Nazionale di Fisica Nucleare, Sezione di Ferrara, 
      via Paradiso 12, I-44100 Ferrara, Italy\\
$^{3}$DAPNIA/SPP, CEA Saclay,
91191 Gif-sur-Yvette, France  \\
$^{4}$Istituto Nazionale di Fisica Nucleare,
 Laboratori Nazionali del Gran Sasso, 
       SS. 16 bis, I-67010 Assergi (AQ),Italy\\
$^{5}$Dipartimento di Fisica dell'Universit\`a di Pisa, 
       P.zza Torricelli 2, I-56100 Pisa, Italy  }

\date{October 1998}
\maketitle                 
\begin{abstract}
The excess of highest energy solar-neutrino events recently observed by 
Superkamiokande can be in principle explained by anomalously high 
$hep$-neutrino flux $\Phi_{\nu}(hep)$. 
Without using SSM calculations, from the solar luminosity
constraint we derive that $\Phi_\nu(hep)/S_{13}$ cannot exceed the SSM
estimate by more than a factor three. If one makes the
additional hypothesis that $hep$ neutrino  production
occurs where the $^3$He concentration is at equilibrium,
helioseismology gives an upper bound which is (less then) two times
the SSM prediction.
We argue that the anomalous $hep$-neutrino flux
of order of that observed by Superkamiokande cannot be explained by
astrophysics, but rather by a large production cross-section.  
\end{abstract}

\section{Introduction}
\label{intro}

In the recent observations of Superkamiokande \cite{REF_SKAM} 
some excess of high energy solar-neutrino events was detected. This excess 
is difficult to interpret as distortion of Boron neutrino spectrum due 
to neutrino oscillations \cite{REF_SKAM,REF_BKS}. It might indicate 
\cite{REF_BAHC} that the $hep$ neutrino flux, $\Phi_{\nu}(hep)$, 
is significantly larger (by a factor $\sim 30$) than the
SSM prediction $\Phi_{\nu}^{SSM}(hep)$.

Apart from $S_{13}$, the zero-energy  astrophysical
$S$-factor of the \mbox{$p+^3He \rightarrow ^4He + e^+ + \nu$}
cross-section, the prediction of the $hep$ neutrino flux in the SSM is rather 
robust. Bahcall and Krastev \cite{REF_BAHC} estimate this flux as:
\begin{equation}
\label{EQ1}
\Phi_{\nu}(hep) = 2.1 (1 +0.03) \left ( \frac{S_{13}}{S_{13,SSM}} \right ) 
\cdot 10^3 cm^{-2} s^{-1}
\end{equation}

We remark that $S_{13}$ is not reliably 
calculated. In the SSM the value $S_{13}^{SSM}=2.3 \cdot 10^{-20}$ keV b is 
used following the most recent calculations by Schiavilla et al \cite{REF_SCH},
though due to complexity of the calculations (see Carlson et al 
\cite{REF_CAR}, Schiavilla et al. \cite{REF_SCH}) the uncertainties are 
rather large: $0.5 <S_{13}/S_{13}^{SSM} < 1.5$, according to Ref.\cite{REF_SCH}. 
In a  short 
review of the calculations \cite{REF_BAHC}, the authors 
conclude that from the first-principle physics it is difficult to 
exclude that the cross-section is an order of magnitude 
larger.

The 3\% error in Eq. (\ref{EQ1}) accounts for the estimated uncertainties
in the solar age, chemical composition, luminosity, radiative opacity,
diffusion rate and in all nuclear quantites, except $S_{13}$.
This small error follows from  the fact that $\Phi_{\nu}(hep)$  depends 
rather weakly on all astrophysical variables such as
temperature $T$, density $\rho$ and the chemical composition. Besides, all
these quantities, except $^3$He abundance, are smooth functions of the 
radial distance $r$ in the solar region where most of $hep$-neutrinos
are produced, $0.1<r/R_{\odot}<0.2$. 
It is hard to conceive that  SSM's uncertainties in $T$, $\rho$ and $X$ 
can result in a considerable change of $\Phi_{\nu}(hep)$. 

The only 
exception is $^3$He abundance, which radial behaviour is not that smooth. 
In fact it increases by an order of magnitude when moving from $r=0.1R_\odot$
to $r=0.2 R_{\odot}$, so that $\Phi_{\nu}(hep)$ is sensitive
to the $^3$He distribution, see Figs.4.2 and 6.1 of \cite{REF_BAH_BOOK}.
This abundance is not limited by helioseismic data 
and in non-standard models it can be, in principle, high in the $hep$-neutrinos 
production zone. 

We have thus analyzed the astrophysical uncertainties in the flux of $hep$
neutrinos in an approach beyond the SSM.
In Section II we 
derive an upper limit for $\Phi_{\nu}(hep)/S_{13}$ directly from the 
solar-luminosity constraint. In Section III we impose the more restrictive 
assumption of {\em local} $^3$He equilibrium in the $hep$-neutrino 
production zone and use the helioseismic constraints.

\section{The solar-luminosity constraint}
\label{solar}

The production rate $Q_{\nu}(hep)$
 for the $hep$ neutrinos and the solar-luminosity constraint 
can be written down as follows
\begin{equation}
\label{EQ2}
Q_{\nu}(hep)=\int dr 4\pi r^2 n_1(r)n_2(r) \lambda_{13}(T(r)),
\end{equation}

\begin{equation}
\label{EQ3} 
\frac{1}{2} \Delta_1  \int dr 4\pi r^2 n_1(r)^2 \lambda_{11}(T(r)) 
+    \frac{1}{2} \Delta_2  \int dr 4\pi r^2 n_3(r)^2 \lambda_{33}(T(r)) \leq 
L_{\odot}
\end{equation}
where $\lambda_{ij}$ are energy-averaged reaction rates between nucleus 
$i$ and $j$,
\begin{equation}
\label{EQ4}
\lambda_{ij}(T)=\int dE dE' f(E,T)f(E',T)(\sigma v)_{ij},
\end{equation}
$f(E,T)$ is the normalized Maxwell distribution function,
$n_i$ is the number density of nuclei with atomic mass
number $i$, $(\sigma v)_{ij}$ is the reaction rate between nuclei $i$ and $j$,
and the two $\Delta$ correspond respectively to the two values of the heat
release:
when a $^3$He nucleus is produced ($3p+e^- \rightarrow ^3He+\nu$), and when
two $^3$He are merged ($^3He+^3He \rightarrow ^4He +2p$),
\begin{mathletters}
\begin{eqnarray}
\label{EQ5a}
  \Delta_1&=& 3 m_p + m_e -m_{^3He} -<E_{\nu}>_{pp} = 6.7 \, {\mbox{MeV}} \\
\label{EQ5b}
  \Delta_3&=& 2 m_{^3He} -  m_{^4He} -2 m_p =        12.9 \, {\mbox{MeV}}
\end{eqnarray}
\end{mathletters}

The temperature dependence of the reaction rates (\ref{EQ4}) can be
parametrized as:
\begin{equation}
\label{EQ6}
  \lambda_{ij}(T) = \lambda_{ij} \left (\frac{T}{T_o} \right )^{\alpha_{ij}}
\end{equation}
where generally speaking $T_0$ is an arbitrary temperature scale. We shall 
fix $T_0$ at the position of maximum of nuclear-energy production in the 
SSM $(T_0=1.336 \cdot 10^7 K)$, which guarantees that that expansion (\ref{EQ6})
is related to a narrow  temperature range. The values of $\lambda_{ij}$ 
and $\alpha_{i,j}$ (this latter rounded to the nearest integer)
 are given in Table \ref{Tab1}. They are calculated using the
values of astrophysical $S$-factors given in Ref.\cite{REF_ALD}  and for 
$T_0= 1.336\cdot 10^7$K.
Note that uncertainties in 
$\lambda_{ij}$ depend only on cross-sections; in particular 
$\lambda_{ij} \propto S_{ij}$, where $S_{ij}$ are astrophysical factors. 

From Eq.(\ref{EQ3}), by using the inequality 
 $1/2(f_1^2+f_2^2) \geq f_1f_2$ and  the
parametrization (\ref{EQ6}) one obtains 
\begin{equation}
\label{EQ7}
\int dr 4\pi r^2 n_1(r)n_3(r)\lambda_{13}(T(r))
\left( \frac{T(r)}{T_0}\right) ^{\alpha} \leq \frac{\lambda_{13}}
{\sqrt{\lambda_{11}\lambda_{33}}\sqrt{\Delta_1\Delta_2}} L_{\odot},
\end{equation}
where $\alpha = (\alpha_{11}+\alpha_{33})/2-\alpha_{13}\simeq 2$

One can see that the integrand in the lhs of Eq.(\ref{EQ7})
is different from that 
of Eq.(\ref{EQ2}) only by a factor $\left( T(r)/T_0\right)^2$. Inequality 
(\ref{EQ7})
further strengthens if this factor is taken as 
$\left( T_{min}/T_0\right) ^2$, 
where $T_{min}$ is the minimum temperature in the $hep$-production 
zone. In the SSM at the temperature $T_{min}=7\cdot 10^6~K$ 
(corresponding to $r/R_{\odot}=0.3$) the probability for a proton to undergo 
a nuclear reaction during the solar age is as small as $0.1\%$. Then from 
Eq.(\ref{EQ7}) one obtains
\begin{equation}
\label{EQ8}
\Phi_{\nu}(hep) < \frac{K_{\odot}}{\sqrt{\Delta_1\Delta_3}}
\frac{\lambda_{13}}{\sqrt{\lambda_{11}\lambda_{33}}}
\left( \frac{T_0}{T_{min}}\right) ^2,
\end{equation} 
where $K_{\odot}=L_{\odot}/(4\pi D^2)$ and $D$ is the distance between 
Sun and Earth. 

Note that the weak inequality (\ref{EQ7}) has turned into a stronger inequality 
(\ref{EQ8}) due to substitution $T(r) \to T_{min}$ in lhs 
of Eq.(\ref{EQ7}), while actually one should use $T(r) \to <T>$.

Using the values from Table 1, one obtains numerically
\begin{equation}
 \Phi_{\nu}(hep) < 6.5   \left ( \frac{S_{13}}{S_{13}^{SSM}} \right ) 
\cdot 10^3 \, {\mbox{ $cm^{-2} \, s^{-1}$}}  \, ,
\end{equation}
a factor three larger than in the SSM calculations, Eq.(\ref{EQ1}).

\section{Local $^3$He equilibrium and Helioseismology}
\label{local}

A more restrictive upper bound can be obtained from
helioseismic constraints, with an additional assumption that $hep$ neutrinos 
are produced
in a region where the $^3$He concentration is at {\em local} equilibrium.
This assumption, which is valid for a wide class of
stellar models, implies:
\begin{equation}
\label{EQ9}    
 \frac{1}{2} n_1^2 \lambda_{11} T(r) = n_3^2 \lambda_{33}T(r) \, .
\end{equation}
Putting $n_3(r)$ from this equation into (\ref{EQ2}) and using 
$n_1(r)=X(r)\rho(r)/m_p$ one obtains
\begin{equation}
\label{EQ11} 
        \Phi_{\nu}(hep) =\frac{\lambda_{13}}{4\pi D^2} 
                        \sqrt{ \frac{\lambda_{11}}{ 2 \lambda_{33}} }
 \frac{1}{m_p^2}  \int dr 4\pi r^2 \rho(r)^2 \left[ X(r) T(r)/T_0\right] ^2 \, .
\end{equation}
In the energy production zone, the equation of state (EOS) for the
solar interior can be approximated, with an accuracy better than
1\% , by the EOS of a fully ionized classical perfect gas:
\begin{equation}
\label{EQ12}
P(r) = \rho(r) T(r) (k_B/m_p) [2X(r) +\frac{3}{4} Y(r) + \frac{1}{2} Z(r)],
\end{equation}
where $P$ denotes the pressure and $k_B$ is the Boltzmann constant. Using  
Eqs. (\ref{EQ11}) and (\ref{EQ12}) one obtains
\begin{equation}
\label{EQ13}
        \Phi_{\nu}(hep) \leq \frac{1}{4\pi D^2}   \frac{1}{4(kT_0)^2} 
         \lambda_{13} \sqrt{ \frac{\lambda_{11}}{ 2 \lambda_{33}} }
                        \int dr 4\pi r^2 P^2(r) \, .
\end{equation}

It is known  \cite{REF_DWZ} that inverting helioseismic data one can
derive the (isothermal) sound speed squared, $u=P/\rho$ and $\rho$
with an accuracy of 1\% or better for all $r/R_\odot$ of interest. 
This implies that also pressure $P$ is known with a comparable
accuracy. Since SSMs are in agreement
with helioseismology, one can use the SSM-calculated pressure $P(r)$ to 
evaluate the integral in Eq.(\ref{EQ13}). It gives
\begin{equation}
\label{EQ14}
 \Phi_{\nu}(hep) < 3.5  \left ( \frac{S_{13}}{S_{13}^{SSM}} \right ) \cdot 10^3 
 \, {\mbox{ $cm^{-2} \, s^{-1}$}}  \, .
\end{equation}
This upper bound is  (less then) two times the SSM prediction.

In fact the agreement is even better. Neglecting $Z$ in Eq.(\ref{EQ12}) and 
using $Y \approx 1-X$ one obtains from Eq.(\ref{EQ12})
\begin{equation}
\label{EQ14b}
X(r)T(r) = \frac{4m_p P(r)}{5k_B\rho(r)} -  \frac{3}{5}T(r)
\end{equation}
The first term on rhs of Eq.(\ref{EQ14b}) is determined by helioseismic measurements and 
thus can be taken as in the SSM. Temperature profile $T(r)$ cannot differ 
from that of the SSM more than by 2 -- 3\%. Then $[X(r)T(r)]^2$, the only 
unknown function in the integral (\ref{EQ11}), can differ from the SSM 
value by a few percent only and so does $\Phi_{\nu}(hep)$.

\section{Conclusions}
We have derived an upper limit on $\Phi_{\nu}(hep)/S_{13}$ directly from 
the solar-luminosity constraint. It is only three times  the SSM 
prediction. If one additionally assumes that $hep$ neutrino  production
occurs in the region where the $^3$He concentration is at local equilibrium,
helioseismology provides a formal
upper bound (less than) two times  the SSM prediction. More realistically, 
in this 
case $\Phi_{\nu}(hep)/S_{13}$ can be only a few percent higher than in 
the SSM.  Our limits can be violated only in very exotic models of 
non-stationary sun with non-stationary transport of $^3$He in the inner 
core from outside. This transport should not be accompanied by any 
noticeable transport of other elements, such as $^1$H or $^4$He, otherwise  the
seismically observed sound speed in the inner core will be affected. We 
doubt that such models can be constructed.

In principle, the $hep$-neutrinos can be resolved  in the high precision 
experiments. We argue that the anomalous $hep$-neutrino flux 
of order of that observed by Superkamiokande cannot be explained by 
astrophysics, but rather by a large production cross-section.

\acknowledgments
We are grateful to  J.N.Bahcall, W.Dziembowski, E. Lisi,
 M.Lissia, M. Spiro and D. Vignaud for interesting 
discussions.
G.F. thanks the DAPNIA of CEA-Saclay  for hospitality in
a stimulating environment.

\begin{table}
\caption[aa]{Parameters of the reaction rates (6)}
\begin{tabular}{llccc}
\hline
& i,j    &    $\lambda_{ij}$      &     $\alpha_{ij}$ &\\
&        &    [$cm^3 s^{-1}$]    &     &\\
\hline
& 1,1    &    8.34  $10^{-44}$   &        4 &\\
& 1,3    &    4.31  $10^{-47}$   &        8 &\\
& 3,3    &    5.88  $10^{-35}$   &       16 &\\
\end{tabular}
\label{Tab1}
\end{table}

\end{document}